\documentclass{article}
\usepackage{mrs2005,epsfig}
\begin{document} 
\title{THE EVOLUTION OF STELLAR MASS IN THE NICMOS UDF AND THE CFHTLS
DEEP FIELDS}

\author{Stephen Gwyn, F.~D.~A.~Hartwick, Anudeep Kanwar, David Schade, Luc Simard}
\affil{University of Victoria, Herzberg Institute of Astrophysics}

\begin{abstract} 
We measure the build-up of the stellar mass of galaxies from $z=6$ to
$z=1$. Using 15 band multicolour imaging data in the NICMOS Ultra Deep
Field we derive photometric redshifts and masses for 796 galaxies down
to $H_{AB}=26.5$.  The derived evolution of the global stellar mass
density of galaxies is consistent with previous star formation rate
density measurements over the observed range of redshifts.  Ongoing
research in the CFHTLS Deep Fields confirms this result at lower
redshifts. Further, if the sample is split by morphological type, a
substantial increase is seen in the number of bulge dominated galaxies
relative to disk-dominated galaxies since $z=1$.
\end{abstract} 
 
\section{Introduction} 
 
For the last ten years, the Lilly-Madau diagram
\cite{cfrssfr,harry,mad1998} has been central to the discussion of
galaxy evolution.  It shows that the star formation rate density
(SFRD) increases with redshift to $z\sim1$ and decreases beyond
that.  In the last year, thanks to the Ultra Deep Field (UDF), several
points have been added at the high-redshift end of the diagram
\cite{bouwens6,bouwens7,bunkerudf,stancdfs}.

While the Lilly-Madau diagram is a useful tool for studying galaxy
evolution, it is subject to some uncertainties, particularly at the
high-redshift end. One source of uncertainty lies in identifying the
high-redshift galaxies: the Balmer break may be confused with the
Lyman break, putting spurious galaxies {\em in} the sample, and
galaxies with heavy extinction may be left {\em out}. Extinction must
also be considered when converting the UV flux into a SFR.  Typically,
a factor of 5 is used for the extinction correction, but the exact
value is imperfectly known.  For these reasons, it would be satisfying
to have some corroboration of the SFR.  This paper presents
measurements of the global stellar mass density (GSMD) based on the
NICMOS UDF.

It is also interesting to find out where and when exactly the mass
builds up. Is it in disk galaxies or ellipticals? Red (older) galaxies
or blue ones?  High density or low density regions?  This paper also
presents preliminary measurements of stellar mass density split
by morphological type in the CFHTLS Deep Fields.

\section{NICMOS Ultra Deep Field: Global Mass Evolution to $z=6$} 

The NICMOS UDF lies within the GOODS South region of the sky and
has been imaged by a large number of telescopes at a variety
of wavelengths. For this project four sources of imaging data for this field
were considered:
\begin{itemize}
\item F110W band ($J$) and F160W ($H$) from NICMOS on HST.
\item F435W ($B$), F606W (somewhere between $V$ and $R$), F775W ($I$)
and F850LP ($Z$).
\item $JHK'$ imaging from ISAAC on VLT.
\item $U'UBVRI$ from the ESO Imaging Survey.
\end{itemize}

The F160W image was used as a reference image. All the other images
were resampled using 
SWarp\footnote{SWarp is available at:  http://terapix.iap.fr/rubrique.php?id\underline{~~}rubrique$=$49}
to match this image. {\tt
SExtractor} \cite{hihi} was run in double-image mode on all the images
in all the bands with the NICMOS F160W ($H$-band) image as the
reference image. Variable elliptical apertures ({\tt SExtractor}'s
{\tt MAG\underline{~~}AUTO}) were used.

Photometric redshifts were calculated for all the objects in the
field. The usual template-fitting, $\chi^2$ minimization method
\cite{ls86a,thesis} was used. The templates were based on the Coleman,
Wu \& Weedman (1980) \cite{cww} spectra. These are supplemented with
the SB2 and SB3 spectra from Kinney et al. (1996) \cite{kin96} as well
as templates interpolated between the main 6 templates. The results
are remarkably good. Comparison with the available secure
spectroscopic redshifts show a $\sigma_z = 0.06\times(1+z)$ spread
with no catastrophic failures. If the less secure spectroscopic
redshifts (including some AGNs) are included, this spread increases to
$\sigma_z = 0.12 (1+z)$, with 2 catastrophic failures for 39
objects. The resampled images and catalogues are available on the 
web\footnote{http://orca.phys.uvic.ca/$\sim$gwyn/MMM/nicmos.html}.

We determined masses for each galaxy with a template fitting process
similar to the photometric redshift method.  The photometry for each
galaxy is converted into an SED and compared to series of templates as
before.  In this case, the redshift of the templates being considered
is held fixed during the $\chi^2$ minimization.

Rather than the empirical Coleman, Wu \& Weedman (1980)\cite{cww} and
Kinney et al. (1996)\cite{kin96} templates, a selection of the PEGASE
2.0\cite{pegase} galaxy spectral evolution models  were used as
templates. The templates were redshifted as before and the IGM
correction was applied. 

The models span the full range of ages from $t=0$ to 14~Gyr. The
metallicity was set to zero (no metals) at $t=0$ in the models. As
each model evolves in time, the metallicity evolves self-consistently.
We added extinction to the models using the reddening curve of Calzetti et al.\cite{calrec}. 
The amount of extinction was varied from $A(V)=0$ to 1.
The Kroupa (1993)\cite{kroupa} initial mass function (IMF) was used
exclusively. The error on the derived masses is a function of apparent
magnitude. It was estimated to be $\sigma_{\rm mass}=0.1$ for
$H_{AB}<25$ and $\sigma_{\rm mass}=0.2$ for $H_{AB}>25$.

The $1/V_a$ method was used to compute mass functions in a series of
redshift bins. The $k$-corrections were computed by interpolation
using the best-fitting template from the photometric redshift
procedure.  To extrapolate beyond the observed range, Schechter 
functions were fit to the binned data. Integrating over these
functions gives the total masses for each redshift bin.

To determine uncertainties on the global mass estimates, a Monte Carlo
method was used.  In the original catalogue of galaxies we added noise
to the measured redshifts and masses according to the error estimates
discussed above.  Further, we simulated the effects of redshift error
on the derived masses by noting the relative shift in luminosity
distance caused by the redshift error, and applying the same shift to
the mass. From this ``noisy'' catalogue we derived mass functions,
integrated over fitted Schechter functions and computed total stellar
masses.  The RMS of the range of total stellar masses derived after
100 actualizations was used for the error bars in Figure 1.

\begin{figure}  
\label{fig:massnic}
\begin{center}
\epsfig{figure=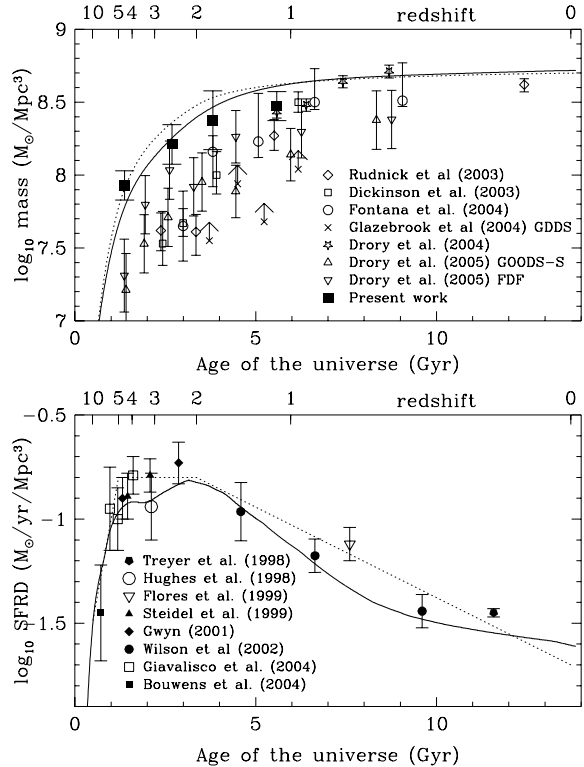,width=8cm}  
\end{center}
\vspace*{0.25cm}  
\caption{Stellar mass evolution. The bottom panel shows the SFR
history of the universe. The points indicate measurements of the SFRD
from the literature.  The top panel shows the stellar mass evolution
from this work along with data from the literature. The GDDS points
are shown as lower limits.  In both panels, the solid line shows the
model of \cite{hartwick2004}.  The dotted line is a arbitrary
parameterization the SFR history (not a model) in the lower panel, and
the integral of this parameterization in the upper panel.  This figure
is based on Figure 3 of Gwyn \& Hartwick (2005).}
\end{figure} 

Figure 1 shows the build up of stars in galaxies as
function of time.  The top panel shows the GSMD, while the bottom
panel shows the SFR.  The top panel can be thought of as $M_\star(t)$,
while the bottom panel can be thought of as its derivative,
$dM_\star(t)/dt$.  This panel is the reverse of the Lilly-Madau
diagram, with cosmic time instead of redshift on the horizontal axis.
The bottom panel shows a number of measurements of the SFR from the
literature
(\cite{treyer1998,hughsubmm,thesis,flores,stei99,wilson,giavalisco2004})
from the compilation of Hartwick (2004)\cite{hartwick2004}.

The upper panel shows our measurements of the GSMD as solid points,
together with a number of measurements from the literature
(\cite{rudnick,dickmass,k20,drory2004,drory2005}) as assorted open
points.  The corrections for the choice of IMF have been applied.

The solid line in both panels of Figure 1 come from Hartwick (2004)
\cite{hartwick2004}, who derived the global star formation history
from observations of the local universe.  Briefly, the model uses the
distribution in metallicity of stars to derive $dM_\star/dZ$ (where
$Z$ is the metallicity) and the age-metallicity relationship for
globular clusters to derive $dZ/dt$ (where $t$ is the age of the
universe).  Combining $dM_\star/dZ$ and $dZ/dt$, one obtains
$dM_\star/dt \equiv$ SFR. This simple model
does a very good job of explaining the star formation history of the
universe, as shown by the agreement between it and the observations in
the lower panel. Using this star formation history as an input to the
PEGASE 2.0 software, we compute a model GSMD.
The result is plotted in the upper panel. It is in excellent agreement
with our GSMD measurements.  Note that this
agreement is not dependent on the details of the Hartwick
model. Almost any description of the star formation history which
agrees with the observed SFRs will, once integrated,
produce good agreement with GSMD. This is
illustrated by the dotted line in Figure 1.  On the lower
panel, this shows an arbitrary, three-segment ``connect-the-dots''
description of the star formation history. The three segments
represent the rise of the SFR at $z<2$, a plateau at $2>z<5$ and the
fall off at $z>5$.  In the upper panel the dotted line shows the results of
integrating (again with PEGASE 2.0) this star formation history.
Again, there is good agreement with our GSMD
calculations.

\section{CFHTLS Deep Fields: Mass Evolution by Morphological Type}

This section presents results from ongoing research in the CFHT Legacy
Survey. The CFHTLS Deep Fields are four pointings of the MegaCam
wide field imager. Each field is 1 square degree and is imaged in the
$ugriz$ filters. The primary scientific goal of the Deep Fields is
supernova cosmology. The fields are reimaged about once every two
weeks and the supernovae are detected by comparing the current image
with the previous image. The total exposure times are currently about
40 hours in the $i$ band.

All the currently available exposures were astrometrically and
photometrically calibrated using the {\tt AstroGwyn} and {\tt
PhotoGwyn} packages and stacked using {\tt Swarp}.  The depth of the
stacked images in AB magnitudes is roughly 26 for the $ugri$ images
and about 24 for the $z$ image.

Photometric redshifts were calculated for every galaxy in the fields.
A comparison with available spectroscopic redshifts from the DEEP2
galaxy survey shows a scatter of about $\sigma=0.08 \times (1+z)$
with less than 1\% catastrophic failures. The stacked
images are available on the web to the Canadian and French
communities\footnote{http://orca.phys.uvic.ca/$\sim$gwyn/cfhtls/index.html}

The CFHTLS Deep Fields do not go as far out in redshift as the NICMOS
UDF does. Also, the lack of infrared data means that care must be
taken when measuring galaxy masses. However, the CFHTLS Deep Fields
sample a far greater volume and consequently contain a much larger
number of galaxies. The total sample can be split into a number of
different subsamples to examine where stellar mass is building
up. Ultimately, the sample will be split by colour (as a proxy for
age) and by local environment (low density {\it vs.} high density) but
for this work the sample was split by morphology.

Bulge-disk decomposition was done for every galaxy brighter than 
$i_{AB}=25$ using the {\tt galfit} \cite{galfit} package. 
The sample was split by bulge-to-total ratio. Galaxies
with B/T$<$0.5 were assigned to the disk-dominated sample;
those with B/T$>$0.5  to the bulge-dominated sample.

In the absence of infrared data, measurements of the stellar mass of
galaxies must be made with caution.  There is a danger that transitory
starburst events will mask the underlying population. In the worst
case scenario, as discussed by Rudnick et al.\cite{rudnick}, this can
lead to over-estimating the stellar mass by 70\%. To avoid this, the
galaxies in each bin (in B/T and redshift) were averaged together. For
each redshift bin, luminosity functions were generated in 5 artificial
bandpasses.  These bandpasses correspond to the original 5 $ugriz$
filters of the CFHTLS survey, but they are blue-shifted so that the
artificial rest-frame bandpass is close to the observed filter. This
minimizes the $k$-corrections.  Integrating over these luminosity
functions produces a coarse SED of the typical galaxy for that
redshift and B/T bin.  Fitting PEGASE2.0 model spectra to this SED
yields a total stellar mass for the bin. By averaging in this way, the
effects of any one starburst occurring in a particular galaxy are
effectively masked.  The only way the mass estimates can be biased is
if most of the galaxies in a given bin have peculiar star-formation
histories.

\begin{figure}  
\label{fig:btmassbw}
\begin{center}
\epsfig{figure=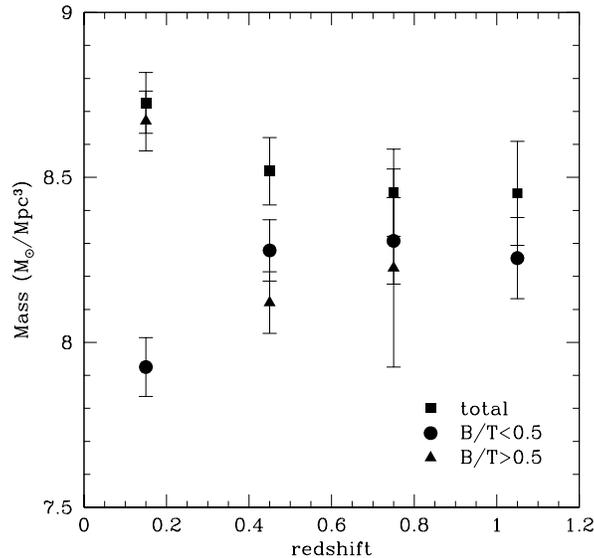,width=8cm}  
\end{center}
\vspace*{0.25cm}  
\caption{Stellar mass evolution for different morphological types:
The squares show the total mass in stars as a function of redshift.
The circles and triangles show the same, but for disk-dominated
and bulge-dominated galaxies respectively.} 
\end{figure} 

The results of this procedure are shown in Figure 2.
The total mass evolution agrees well with the GSMD
from Figure 1. The figure also shows that the mass
density of bulge-dominated galaxies is increasing rapidly as a
function of time (note that there are almost no bulge-dominated galaxies at
$z>1$; this point is off the graph) while the mass in disk-dominated
galaxies is dropping off slightly with increasing time/decreasing
redshift.  This result is similar to that of Bundy et
al.\cite{kbundy}.  They split their sample into ellipticals, disks and
peculiars using traditional visual classification. They found that the
mass in elliptical galaxies increased with time at the expense of disk
and peculiar galaxies. Their differential evolution is not as extreme
as the results presented here, probably because the $B/T>0.5$ cut used
to select bulge-dominated galaxies also selected some earlier-type
disk systems. The combination of the present work and Bundy et
al.\cite{kbundy} argues that there is a strong trend 
toward more massive bulges in galaxies since $z=1$, with a less extreme
trend towards complete disruption of their disks.

\section{Summary} 
\begin{itemize}
\item Work on the NICMOS UDF shows that the global stellar mass build-up
is consistent with what one would predict from measurements of
the SFR.
\item Work on the CFHTLS Deep Fields shows that the mass in
disk-dominated galaxies decreases slightly over time while the mass in
bulge-dominated galaxies increases greatly over time.
\end{itemize}
 
\acknowledgements{S.D.J.G. was supported partially by a discovery
grant from NSERC and from an NSERC CRO grant which supports Canadian
participation in the CFHT Legacy Survey.  }

\vfill 
\end{document}